\documentstyle[12pt]{article}
\title{\Large\bf BEKENSTEIN MODEL AND THE TIME
VARIATION OF THE STRONG COUPLING CONSTANT}

\author
{ \it \bf  N. Chamoun$^{1,2}$, S. J. Landau$^{3}$\thanks{fellow of
CONICET} , H. Vucetich$^{1,3}$\thanks{member of CONICET} ,
\\ \small$^1$ Departamento de F\'{\i}sica, Universidad Nacional de
La Plata,\\ \small c.c. 67, 1900 La Plata, Argentina. \\
\small$^2$ Department of Physics, Higher Institute for Applied
Sciences and Technology,\\ \small P.O. Box 31983, Damascus, Syria.
\\ \small$^3$ Observatorio Astron\'{o}mico, Universidad Nacional
de La Plata,\\ \small Paseo del Bosque S$/$N, CP 1900 La Plata,
Argentina.}

\date{}

\begin{document}
\maketitle
\begin{abstract}
We propose to generalize Bekenstein model for the time variation
of the fine structure ``constant" $\alpha_{em}$ to QCD strong
coupling constant $\alpha_S$. We find that, except for a ``fine
tuned'' choice of the free parameters, the extension can not be
performed trivially without being in conflict with experimental
constraints and this rules out $\alpha_S$ variability. This is due
largely to the huge numerical value of the QCD vacuum gluon
condensate when compared to the matter density of the universe.
\end{abstract}



\section{Introduction}
\label{Introduction}
 The time variation of fundamental constants may
provide a
 connection between cosmology and particle physics.
Early
 suggestions can be traced to Dirac \cite{dirac37}
long ago, but
 many proposals leading to time-varying constants
were discussed
 far afterwards. These can be classified to either
 ``phenomenological" models
 \cite{vucetich90,vucetich91,albrecht99}, models
providing a
 natural theoretical framework in terms of
higher-dimensional
 theories like Kaluza-Klein
\cite{marciano84,vucetich89} and string
 theories \cite{wu86,maeda88} or models based on first principles such as
 the Bekenstein's for $\alpha_{em}$ variability
 \cite{bekenstein82}. Besides, spatial variations of the fine structure within Beckenstein's models have also been considered \cite{barrow99}

Actually, Bekenstein model for electromagnetism
 is very attractive because it is based on very general assumptions:
 covariance, gauge invariance, causality and time-reversal
 invariance of electromagnetism as well as the idea that the
 Planck-Wheeler length ($10^{-33} \, \rm{cm}$) is the shortest scale allowable
 in any physical theory. 
The very generality of its assumptions
 guarantee the applicability of the scheme to other ``gauge"
 interactions such as the strong forces. Besides, it introduces a useful simplifying assumption; namely, that the gravitational sector is unaffected by the scalar field introduced to vary the coupling constant . However, it is interesting to explore first this simplified model, before a similar exploration of more general theories.

In fact, a grand unified
 description of electromagnetism and QCD will, presumably, predict linked time
 variation of low energy fundamental constants
 \cite{marciano84,Drinkwater98,Olive95}. Nevertheless, one can anticipate 
that the QCD vacuum contribution, as we shall see in section 3, will 
be several orders of magnitud larger than the contribution coming
 from $\alpha_{em}$ variability \cite{landau00}, so that we can assume, in the presence of a variable
 unifying coupling constant, that the strong forces separate
 cleanly from other interactions and so study them alone. To date we know of no
 ``gauge-principled'' analysis for the variability of the strong
 coupling constant and the object of this letter is to provide just
 such a study. In particular we will apply Bekenstein
scheme to QCD since its assumptions are still valid for the strong
interactions and find, contrary to the case of electromagnetism,
that it is the vacuum, when compared to matter, which plays the
dominant role as a source of variability.

\section{Analysis}
\label{analysis}

Our starting point is the QCD Lagrangian with a varying coupling
``constant"
\begin{eqnarray}
L_{QCD}&=& L_{\epsilon}+L_g+L_m \nonumber\\ &=&L_{\epsilon}
-\frac{1}{2}Tr(G^{\mu\nu}G_{\mu\nu}) + \sum_f \bar{\psi}^{(f)}(i
\gamma^\mu \partial_\mu -M_f +g_0\epsilon(x) A_\mu \gamma^\mu )
\psi^{(f)}
\end{eqnarray}
where $G^{\mu\nu}=G^{\mu\nu}_a t^a$, $A^\mu=A^\mu_a t^a$ ,
$[t^a,t^b]=i f^{abc} t^c$ , $M_f$ is the $f$-flavour quark mass and
where, follwing \cite{bekenstein82}, we introduced a classical
scalar gauge-invariant and dimensionless field $\epsilon(x)$. The
varying coupling constant is given by $g(x)=g_0\epsilon(x)$ where
$g_0$ is a constant and we require our theory governing $\epsilon$
to be scale invariant.

In order that the action be invariant under
$\psi\rightarrow\psi^\prime = U\psi$ ($U=e^{-it^a \theta_a(x)}$),
we find that $\epsilon A_\mu\rightarrow\epsilon A_\mu^\prime =
U\epsilon A_\mu U^{-1} -\frac{i}{g_0}(\partial_\mu U) U^{-1}$
while the gluon tensor field, transforming like $G\rightarrow
G^\prime=UGU^{-1}$, should be given by
\begin{eqnarray}
\label{gluon equation} G^a_{\mu\nu} &=& \frac{1}{\epsilon}
[\partial_\mu(\epsilon A^a_\nu)-\partial_\nu(\epsilon
A^a_\mu)+g_0\epsilon^2f^{abc}A^b_\mu A^c_\nu]
\end{eqnarray}
Similar to the electromagnetic case, it is the time reversal
invariance which excludes the $G^*G$ term from the free gluon
action while, concerning the dynamics of $\epsilon$, the same
arguments in \cite{bekenstein82} apply so we take
\begin{eqnarray}
L_\epsilon &=& -\frac{1}{2} \frac{\hbar c
}{l^2}\frac{\epsilon_{,\mu}\epsilon^{,\mu}}{\epsilon^2}
\end{eqnarray}
where we merely require the scale length $l$ to be no shorter than
the Plank-Wheeler length $L_{P}=\sqrt{\frac{\hbar G}{c^3}}$.

Writing the Euler-Lagrange equations for the total action $S=\int
L_{QCD}\sqrt{-g}d^4x$, first with respect to $A^\mu$, we find
\begin{eqnarray}
\label{equation1}
\left(\frac{G^{\mu\nu}_a}{\epsilon}\right)_{;\mu}-g_0f^{abc}G^{\mu\nu}_b
A^c_\mu+\sum_f g_0\bar{\psi}t^a \gamma^\nu\psi &=&0
\end{eqnarray}
then with respect to $\epsilon$, we get
\begin{eqnarray}
\label{equation2} \Box\ln\epsilon&=& \frac{l^2}{\hbar c}\left[
(G_a^{\mu\nu} A^a_\mu)_{;\nu}+g_0\frac{\epsilon}{2}f^{abc}
G^{\mu\nu}_a A^b_\mu A^c_\nu \right. \nonumber\\ &&\left.
+\sum_f\bar{\psi} \frac{\partial M_f}{\partial \epsilon} \epsilon
\psi - \sum_f g_0 \epsilon \bar{\psi} A_\mu \gamma^\mu \psi
\right]
\end{eqnarray}
Substituting (\ref{equation1}) in (\ref{equation2}) we obtain
\begin{eqnarray}
\Box\ln\epsilon&=&\frac{l^2}{\hbar c}\left[\sum_f\bar{\psi}
\frac{\partial M_f}{\partial \epsilon} \epsilon \psi
-\frac{1}{2}G_a^{\mu\nu}G^a_{\mu\nu} \right]
\end{eqnarray}

This equation is analogous to the electromagnetic case but while,
in electromagnetism, only matter acted as a source for both terms,
by contrast, in QCD we can drop the first term working in the
chiral limit $M_f \approx 0$ and, more importantly, we should include
the vacuum contribution to the energy density in the second term.
Approximating the mass in the universe by free nucleons we can
write
\begin{eqnarray}
\Box\ln\epsilon&=&\frac{l^2}{\hbar c}\left[-\frac{1}{2} \langle
0\mid G_a^{\mu\nu}G^a_{\mu\nu} \mid 0\rangle -
\sum_{nucleons}\frac{ \langle N\mid Tr(G^{\mu\nu}G_{\mu\nu}) \mid
N\rangle}{\langle N \mid N \rangle} \right] \nonumber
\\ &=&
\frac{-l^2}{2\hbar c}\left[ \langle 0\mid G_a^{\mu\nu}G^a_{\mu\nu}
\mid 0\rangle + \rho_m A^{(2)}_g \right]
\end{eqnarray}
where $\rho_m$ is the matter density of the universe and where the
matrix element $A^{(2)}_g$ is defined for the twist-2 operator by
\begin{eqnarray}
\langle N\mid Tr(G^{\mu_1\nu}G_{\mu_2\nu}) - traces \mid N\rangle
&=& A^{(2)}_g (p^{\mu_1}p^{\mu_2}- traces )
\end{eqnarray}
and has the physical meaning of the part of nucleon momentum
carried by gluons.

Assuming homogeneity and isotropy for an expanding universe, and
so considering only temporal variations for $\alpha_S$, we get
\begin{eqnarray}
\label{cosequation}
(a^3 \frac{\dot{\epsilon}}{\epsilon})^.
&=&\frac{a^3(t) l^2}{2\hbar c}c^2\left[ \langle G^2 \rangle +
A^{(2)}_g \rho_m \right]
\end{eqnarray}
where $a(t)$ is the expansion scale factor.
Let us consider models
where the scale factor of the Universe behaves as in a flat
Robertson-Walker space-time. Thus, for models with cosmological constant, we have 
\begin{eqnarray}
\label{a2}
a(t)&=&a(t_0)\left( \frac{\Omega_m}{\Omega_\Lambda}
\right)^\frac{1}{3} \left[ \sinh \frac{3}{2}
\Omega_\Lambda^\frac{1}{2} H_0 t \right] ^\frac{2}{3}
\end{eqnarray}
where $\Omega_{m,\Lambda}$ is the cosmological density parameter
corresponding to the mass and the cosmological constant
respectively  and $\Omega_m + \Omega_{\Lambda}=1$ for flat universes. Models without cosmological constant are obtained in the limit $\Omega_{\Lambda}\rightarrow 0 $, where the scale factor behaves as:
\begin{eqnarray}
\label{a3}
a(t)&=&a(t_0)\left( \frac{t}{t_0} \right)^\frac{2}{3}
\end{eqnarray}

\section{Results and Conclusion}
The matrix element $A^{(2)}_g$ can not be computed perturbatively
and experiments give it the value of $0.48$ \cite{yndurain99}
while for the other non-perturbative quantity: the gluon
condensate, we can estimate it by QCD sum rules method
\cite{shifman92} \[\langle 0 \mid \frac{\alpha_S}{\pi} G^2 \mid 0
\rangle \sim (0.012 \pm 0.004)\,\rm{ GeV^4}\] where the operator $G^2$ is
renormalized at the ``natural'' scale $1 \,\rm{GeV}$ corresponding to the
matching condition of the sum rules to avoid the appearance of
large radiative corrections. In order to evaluate $\langle G^2
\rangle$ we take $\Lambda_{QCD} = 125 \pm 25 \,\rm{MeV}$ consistent with
the range of values used in QCD sum rules \cite{ross96} implying,
to leading log, $\alpha_S(1\,\rm{GeV})=0.336\pm0.0323$ and so we find
\[\langle  G^2  \rangle \sim
(0.112 \pm 0.048)\,\rm{GeV^4}\]
Now since, in our model, $\rho_m =
\frac{3 H_0^2 c^2}{8\pi G} \left( \frac{a(t_0)}{a(t)} \right) ^3$
with $\rho_c = \frac{3 H_0^2 c^2}{8\pi G} \sim 10^{-47} $ GeV$^4$ we
find, here, a strikingly exotic predominance of the QCD vacuum over
matter and we can neglect the mass term altogether in equation
(\ref{cosequation}). By integration we find then for models with
cosmological constant:
\begin{eqnarray}
\label{eps}
\frac{\dot{\epsilon}}{\epsilon}&=&
(\frac{l}{L_P})^2
\frac{\langle G^2 \rangle}{\rho_c} \frac{3 H_0^2}{16\pi}
\frac{\Omega_m}{\Omega_{\Lambda}}\frac{a_0^3}{a^3(t)} \times
\nonumber\\&& \left[
-\frac{(t-t_c)}{2}+\frac{\sinh(3\sqrt{\Omega_\Lambda}H_0 t)}{6 H_0
\sqrt{\Omega_\Lambda}} -\frac{\sinh(3\sqrt{\Omega_\Lambda}H_0
t_c)}{6 H_0 \sqrt{\Omega_\Lambda}}\right]
\end{eqnarray}
where $t_c$ is an unknown free parameter.
In the limit of zero cosmological costant, the variation 
of $\epsilon$ behaves as:
\begin{eqnarray}
\label{eps2}
\frac{\dot{\epsilon}}{\epsilon}&=&
(\frac{l}{L_P})^2
\frac{\langle G^2 \rangle}{\rho_c}\frac{H_0^2}{16 \pi} 
\left( t - \frac{t_c^3}{t^2}\right)
\end{eqnarray}
In order to evaluate equations (\ref{eps}) and (\ref{eps2}) for today we
need laboratory bounds on the variation of the strong coupling
constant. Here, we can use a large number of data from various high
energy processes ordered by increasing energy scale of the
measurement as follows: $\tau$ decay, $GLS$ sum rule, $Q\overline{Q}$
lattice, deep inelastic scattering, $R(e^{+},e^{-})$, $P_t(w)$,
$e^{+}e^{-}$ event shape and $Z$ width, giving in all a weighted
average to $\Lambda_{QCD}$ equal to $195\pm65$ MeV in the year 1994
\cite{databook94} and $208\pm25$ MeV in the year 1999
\cite{data99}. This allows us to take, up to leading log terms, the
laboratory bound $\mid\frac{\dot{\epsilon}}{\epsilon}\mid _{today} =
\frac{1}{2}\mid\frac{\dot{\alpha_S}}{\alpha_S}\mid < 4. 10^{-2}
\,\rm{yr}^{-1}$. 

If we assume, plausibly, $t_c$ of the order of $t_0 \sim 10^{10}$
$\,\rm{yr}$ $\sim H_0^{-1}$ and take 
$\Omega_\Lambda$ in the interval $[0.25,0.75]$ with $\Omega_m=1-\Omega_\Lambda$ for models with cosmological constant (equation \ref{a2}) or use equation (\ref{a3}) for models without cosmological constant, we find, pushing $l$ down to near its
minimum allowable value $\frac{l}{L_P}\sim 1$, a constraint on
$\mid t_0 - t_c \mid < 10^{-25} yr$ which is highly strange
barring a ``fine tuning" situation.

For the purpose of refining the analysis, let us substitute
equation (\ref{a2}) into (\ref{eps}) and integrate with
$\epsilon(t_0)=1$ to get the following expression for the
variation of $\alpha_{s}$ for models with cosmological constant

\begin{eqnarray}
\label{variation} \frac{\Delta\alpha_S}{\alpha_S}&=&
\left(\frac{l}{L_P}\right)^2 \frac{\langle G^2 \rangle}{\rho_c}
\frac{1}{12 \pi \Omega_{\Lambda}} \left[ x \coth{x} - x_0
\coth{x_0} - x_c \coth{x} + x_c \coth{x_0}\right. \nonumber
\\&& \left.  + \sinh{x_c}\cosh{x_c}\coth{x} -
\sinh{x_c} \cosh{x_c} \coth{x_0} \right]
\end{eqnarray}
where $ x = \frac{3}{2} \sqrt{\Omega_\Lambda} H_0 t $ and
$x_0$($x_c$) is $x$ evaluated at $t_0$($t_c$).
In the limit of zero c
cosmological constant, we find using equation (\ref{eps2})

\begin{eqnarray}
\label{variation2} \frac{\Delta\alpha_S}{\alpha_S}&=&
\left(\frac{l}{L_P}\right)^2 \frac{\langle G^2 \rangle}{\rho_c}
\frac{H_0^2}{8 \pi} \left(\frac{t^2}{2}-\frac{t_0^2}{2}+\frac{t_c^3}{t}-\frac{t_c^3}{t_0}\right)
\end{eqnarray}

Now, we can use astronomical and geophysical data giving bounds on
the variation of $\alpha_S$ ranging over longer periods of time. In
fact, the authors of \cite{vucetich90} have derived a relation
between the shift in the half-life of three long lived $\beta $
decayers ($^{187}\rm{Re}, ^{40}\rm{K}$ and $ ^{87}\rm{Rb}$),
measured in laboratory or by comparison with the age of meteorites,
and a possible temporal variation of the fundamental constants
$\alpha_{em} ,\Lambda _{QCD}$ and $G_F$. In this work we attribute
the change uniquely to $\Lambda _{QCD}$ and so we get a bound for
the variation of $\alpha_S$ at the age of the meteorites compared to
its value now $\frac{\Delta\alpha_S}{\alpha_S}=(0\pm 2.1
\times 10^{-4})$.

One of the most stringent limits on the time variation of fundamental
constants follows from an analysis of isotope ratios of
$^{149}\rm{Sm}/^{147}\rm{Sm}$ in the natural uranium fission reactor that took
place $1.8\times 10^9$ yr ago at the present day site of the Oklo mine
in Gabon, Africa \cite{Schlyakter76,Damour96}. Sisterna and Vucetich 
\cite{vucetich90} have shown that the
information about the time variation of the strong interaction
parameter $\Lambda_{QCD}$ is very small when compared to the other
fundamental constants. This is because in the chiral limit, the time variation of any strong energy difference takes the form $\frac{\dot{\Delta E}}{\Delta E}= \frac{\dot{\Lambda}}{\Lambda}$; and the limits on the shift of the neutron capture resonance are not very accurate. (See also the discussion in \cite{Damour96}.Thus, we will not include the Oklo limits in our
set of data.

On the other hand, quasar absorption systems present ideal
laboratories to test the temporal variation of the fundamental
constants. The continuum spectrum of a quasar was formed at an epoch
corresponding to the redshift $z$ of the main emission details with
the relation $ \lambda _{obs}=\lambda _{lab}\left( 1+z\right)
$. Knowing that the ratio of frequencies of the hyperfine 21 cm
absorption transition of neutral hydrogen to an optical resonance
transition is proportional to $x=\alpha_{em} ^2g_p\frac{m_e}{m_p}$
where $g_p$ is the proton $g$ factor and $m_p$ is its mass, we can
translate a change in $x$ into a difference between the measured
redshifts of the 21 cm and the optical absorption as follows:
\begin{equation}
\frac{\Delta x}{x} = \frac{z_{opt}-z_{21}}{\left( 1+z\right) }
\end{equation}
Thus, combining the measurements of optical and radio redshifts,
one can obtain bounds on $x$: $\frac{\Delta x}x=(0.7\pm 1.1) \times
10^{-5}$ at $z=1.776$ \cite{CyS}, $\frac{\Delta x}x=(0\pm 1.2) \times
10^{-4}$ at $ z=0.69$ \cite{WolfeyDavis} and $\frac{\Delta x}x=(0\pm
2.8) \times 10^{-4}$ at $z=0.52$ \cite{WolfeyBrown}. 

Evidence for the time variation of the fine structure constant has
been claimed by Webb et al \cite{Webb99} and further results posted recently \cite{Murphy00,Webb00}. However, this detection does not suggest a power law fit.
Moreover, it is found in \cite{livio98} that Webb's results
are in conflict with the equivalence principle when examined
within Bekenstein model for $\alpha_{em}$ variability which is in agreement with the conclusion of \cite{landau00} examining all
available experimental bounds in the context of typical theories
predicting time variation of the fundamental constants. For the purposes of this paper,the
bounds on $x$-variations will be assumed to come solely from the change
of $m_p$ proportional to $\Lambda_{QCD}$ and, thus, infer bounds on
$\alpha_S$-variations. Moreover, observations of molecular hydrogen in
quasar absorption systems can be used to set bounds on the evolution
of $\mu=\frac{m_e}{m_p}$ throughout cosmological time scales:
$\frac{\Delta\mu}{\mu} < 2 \times 10^{-4} $ at $z=2.811$
\cite{pothekin98}, and this, in turn, would imply a bound on
$\alpha_S$-variation under a similar assumption to that for the other
quasar data. Bounds on the mass ratio have also been examined \cite{Pagel83,Varshalovich96} using comparison of 21-cm
hyperfine hydrogen and molecular rotatinal absorption lines. However, it is shown that this comparison
actually constraints $\alpha_{em} g_p$ \cite{Drinkwater98}.

Taking the data described above, with more ``reasonable'' choices for $t_c$, we have performed a statistical analysis using MINUIT with
$H_0=6.64 \times 10^{-11}\,\rm{yr^{-1}}$ and obtained the results in the table \ref{tablaB}
for the free parameter in the model $\frac{l}{L_p}$ with
99 $\%$ of confidence level. Furthermore, we could check that, as long as $t_c$ is far from the unnatural ``fine
tuning" situation, we
always get $l$ shorter than $L_P$ by too many orders of magnitude in
clear conflict with the postulates adopted in the framework. While
integrating, we have assumed the value of the gluon condensate $<G^2>$
constant whereas it might vary over cosmological time scales. However,
we do not expect such variations to be large enough to change the
conclusion above. Even though length scales shorter than the
Planck-Wheeler length $L_P$ might enter physics in the context of
``new" theories, namely superstrings, it is extremely unlikely that
our tiny $l$ could be accommodated so as to recover the axioms of the
Bekenstein model. Since the assumptions of this model are reasonable,
comparison with experiments, excludes any direct
generalization of the model to QCD and consequently rules out
$\alpha_S$ variability in accordance with the strong principle of
equivalence. Although it is possible that more general schemes, involving more realistic assumptions on the structure of the $\epsilon$ field or its interactions with gravitation, may resist comparison with the experiment, the present study shows that it will be difficult to build them from first principles without some fine tunning mechanism.

%

\begin{table}[tbp]
\caption{{} Constraints on $\frac{l}{L_p} $ using the full data set of  bounds on $\frac{\Delta\alpha_s}{\alpha_s}$  for flat models ($\Omega_\Lambda + \Omega_m=1 $) with different cosmological constant.$ t_c=\gamma t_0$}
\label{tablaB}
\begin{center}
\begin{tabular}{lllllll}
\hline
$\Omega _\Lambda $ & & & $ \gamma$ & & & $\frac{l}{L_p}$ \\ 
\hline
 $0.75$ & & & $ 0.3$ & & & $ (0 \pm 5) \times 10^{-25} $ \\ $0.75$ & & & $ 0.7 $ & & & $ (1 \pm 4) \times 10^{-25} $ \\  $0.5$ & & & $0.3$ & & & $ (0  \pm 6) \times 10^{-25}  $  \\  $0.5$ & & & $0.7$ & & & $ (2 \pm 7) \times 10^{-25} $  \\ $0.25$ & & & $0.3$ & & & $ (0 \pm 7)\times 10^{-25} $ \\ $ 0.25 $ & & & $ 0.7 $ & & & $ (4 \pm 8)\times  10^{-25} $\\ $ 0 $ & & & $ 0.3 $ & & & $ (0 \pm 7)\times 10^{-25} $\\ $ 0 $ & & & $ 0.7 $ & & & $ (0 \pm 2) \times 10^{-24} $%
\end{tabular}
\end{center}
\end{table}

\section*{Acknowledgements}
This work was supported in part by CONICET, Argentina. N. C.
recognizes economic support from the Third World Academy of
Science (TWAS).

\end{document}